\documentclass[a4paper,11pt]{article}

\usepackage{jheppub} 

\usepackage{amsmath,amssymb}
\usepackage{listings, color}
\usepackage{hyperref}

\makeatletter
\newcommand{\thickhline}{%
	\noalign {\ifnum 0=`}\fi \hrule height 1pt
	\futurelet \reserved@a \@xhline
}

\newcommand{\E}{\mathcal{E}}
\newcommand{\C}{\mathcal{C}}
\newcommand{\Dl}[1]{\mathchoice{\ffrac{\partial}{\partial #1}}{\frac{\partial}{\partial
      #1}}{\ffrac{\partial}{\partial #1}}{\ffrac{\partial}{\partial #1}}}
\usepackage[mathscr]{eucal}
\usepackage{amsmath,amssymb,amsthm,mathtools}
\usepackage{times}
\usepackage{hyperref}

\renewcommand{\tilde}{\widetilde}
\renewcommand{\hat}{\widehat}

\newcommand{\bref}[1]{\textbf{\ref{#1}}}

\newcommand{\p}[1]{|#1|}
\newcommand{\gh}[1]{\mathrm{gh}(#1)}

\renewcommand{\d}{\partial}

\newcommand{\tensor}{\otimes}

\renewcommand{\geq}{\,{\geqslant}\,}
\renewcommand{\leq}{\,{\leqslant}\,}

\newcommand{\inner}[2]{\langle #1{,}\,#2\rangle}
\newcommand{\binner}[2]{%
  {\langle}\kern-4.15pt{\langle}#1{,}\,#2{\rangle}\kern-4.15pt{\rangle}}
\newcommand{\commut}[2]{[#1{,}\,#2]}

\newcommand{\half}{\mathchoice{%
    \ffrac{1}{2}}{\frac{1}{2}}{\frac{1}{2}}{\frac{1}{2}}}

\newcommand{\ffrac}[2]{\raisebox{.5pt}%
  {\footnotesize$\displaystyle\frac{#1}{#2}$}\kern1pt}

\newcommand{\Liealg}{\mathfrak} 
\newcommand{\algg}{\Liealg{g}}

\newcommand{\algA}{\mathcal{A}}

\newcommand{\fZ}{\mathbb{Z}}

\def\bbeta{{\boldsymbol{\beta}}}

\newcommand{\bA}{\mathbf{A}}
\def\cD{\mathcal{D}}
 \def\cE{\mathcal{E}}
 \def\cH{\mathcal{H}}
 \def\cI{\mathcal{I}}
 \def\cU{\mathcal{U}}

\def\tr{{\rm Tr}}

\def\BG-Poincare{Barnich:2009jy}
\def\Fedosov-book{Fedosov:1996fu}

\begin{document}
\begin{flushright}
	LMU-ASC 16/21\\
\end{flushright}

\title{\boldmath 
A toy model for background independent string field theory
}


\author[a,b]{Maxim Grigoriev,}
\author[c]{Adiel Meyer,}
\author[d]{and Ivo Sachs}


\affiliation[a]{Lebedev Institute of Physics,
  Leninsky ave. 53, 119991 Moscow, Russia }
\affiliation[b]{Institute for Theoretical and Mathematical Physics,
\protect\\
  Lomonosov Moscow State University, 119991 Moscow, Russia}
\affiliation[c]{Department of Natural Sciences, The Open University of Israel, \\ PO Box 808, Ra’anana 43537, Israel}
\affiliation[d]{Arnold-Sommerfeld-Center for Theoretical Physics, Ludwig-Maximilians-Universit\"at, M\"unchen,\\ Theresienstr. 37, D-80333 M\"unchen, Germany}

\emailAdd{grig@lpi.ru}
\emailAdd{adielmb@gmail.com}
\emailAdd{ivo.sachs@physik.lmu.de}


\abstract{
We study gauge theories of background fields associated to BRST quantized spinning particle models and identify background-independent algebraic structures which allow to systematically reduce the spectrum of fields and subject some of them to dynamical equations of motion. More specifically, we construct a manifestly background-independent extension of the model based on $N=2$ spinning particle. The resulting system describes an on-shell spin-1 field coupled to off-shell background fields including metric and dilaton. Tensoring with a given Lie algebra results in a non-abelian extension of the model.
}

\maketitle
\flushbottom

\section{Introduction}
Covariant string field theory \cite{Witten:1986cc,Zwiebach:1993ie} is, by construction, a Batalin-Vilkovisky (BV) formulation of some gauge system. While the origins of the underlying gauge redundancies can easily be traced to (super) reparameterization invariance on the world sheet, their space-time interpretation are still to a large extent obscure. This is in contrast with the BV formulation of the usual gauge theories where the starting point is a dynamical system whose redundancies and equations of motion have a manifest space-time interpretation.

On the other hand, the construction of string field theory usually  requires  some choice of background metric\footnote{See however, \cite{Witten:1992qy,Shatashvili:1993ps} for a proposal of a manifestly background independent formulation of open string field theory.} in order to quantize the world-sheet theory,  although space-time itself is not a background independent concept in string theory. While a perturbative construction around a given background is quite sufficient to address questions like mass renormalization and background shifts {\cite{Sen:2015uoa}} a global description becomes important to understand off-shell symmetries or to look for disconnected critical points describing classical solutions of string theory. Still, in some situations the perturbative formulation may already be sufficient to find non-trivial vacua. This is the case, for instance in Witten's open bosonic string field theory whose equation of motion describe the tachyon vacuum \cite{Schnabl:2005gv} which is not continuously connected to the trivial vacuum. Open string field theory also contains information about closed string equations of motion. Indeed considering open string field theory, on a continuous deformation of the closed string background one finds that classical consistency of open string field theory implies the perturbative equations of motion for the closed string \cite{Moeller:2010mh}. The global structure of closed string field theory, on the other hand, does not appear to be described by the available formulations of open or closed string field theory.

The aim of this work is  to analyze background independent structures of a gauge field theory arising from the operator algebra of the quantized spinning particle seen as a toy model of the quantized string.  More precisely, the point particle limit of string field theory, where the world sheet reduces to a world line. In this model it is possible to obtain a background independent  parametrization of the BV-configuration space starting from the graded Lie 
operator (super)algebra arising from the quantization of the (spinning) world line. In so doing the BRST operator of the model is understood as a generating function for background fields while the nilpotency condition and the natural equivalence are understood as equations of motion and gauge transformations respectively~\cite{Horowitz:1986dta,Grigoriev:2006tt,Dai:2008bh,Bekaert:2013zya}.

In fact, it turns out that even this smaller algebra is typically too rigid to admit nilpotent elements other than that corresponding to the off-shell background fields describing the underlying geometry and hence does not result in dynamical gauge theories.\footnote{These background field can be made dynamical by supplementing them with the induced action~\cite{Segal:2002gd,Tseytlin:2002gz,Bekaert:2010ky} or reinterpreting them as the boundary values of the on-shell fields in the bulk in the context of AdS/CFT correspondence~\cite{Bekaert:2013zya,Bekaert:2017bpy,Grigoriev:2018wrx}}
However, by relaxing the nilpotency condition to hold only on some subspace of the representation space the spectrum naturally extends by genuine on-shell fields. 
Furthermore, the equations imposed by nilpotency on these new fields have a familiar physical interpretation of the
fully non-linear Yang-Mills equations \cite{Dai:2008bh} for $N=2$ spinning particle, Einstein equations \cite{Bonezzi:2018box} for $N=4$ spinning particle or Supergravity equations of motion \cite{Bonezzi:2010jr}. Equations of motion for background fields in the ambitwistor string arise in an algebraically analogous way \cite{Adamo:2014wea}.

These subspaces of the representation space have a simple interpretation in the world line theory. They correspond to gauging subgroups of the R-symmetry group that come with any extended world line supersymmetry. Some of these gaugings also have an interpretation in string theory. They correspond to level matching, level truncation and partity projection on the world sheet. In more algebraic terms the theory is determined by a subalgebra of the graded operator algebra that preserves the subspace and factorized over the left-right ideal of operators that act trivially on the latter. Although this is performed with the graded algebra involving BFV-BRST ghosts, algebraically this is just a usual constraint reduction or a version of quantum Hamiltonian reduction.

The complete characterization of the spectrum of the gauge theory associated to the spinning particle  amounts to working out the operator BRST cohomology in the reduced algebra. This is a tall order, even for the world line, since one expects to have an infinite number of cohomology classes, corresponding to background fields of all spins. Here we will reduce this problem by introducing a filtration with an additional degree that is related to the dimension of elements in the operator algebra. This allows us to disregard the background of higher spins in an invariant way and work out the operator BRST cohomology for some given filtrations for the spinning particle. In particular, for $\mathcal{N}=2$ at filtration $0$ the theory is completely characterised by an on-shell Maxwell field coupled to an unconstrained dilaton field. Tensoring with the gauge Lie algebra results in the non-abelian extension of this model. Gravity arises at filtration $1$ but equations of motion for the latter, as well as for the dilaton only arise in the case of  $\mathcal{N}=4$ worldline SUSY.

Despite being able to completely describe the spectrum only in certain filtrations, 
the theory is perfectly well-defined and is background  independent. More specifically, to define the BV-extended field space and the BRST differential (which in turn determines equations of motion, gauge generators etc.) we do not need to assume any background fields to be predefined in space-time. At least locally, the operator algebra as well as the constraint, determining the reduction, require for their construction only the canonical symplectic form on the contangent bundle to define the quantization,  and the ghost grading.
The subtle point, however, is that the gravity background, contained in the degree 1 component of the algebra, is off-shell i.e. the nilpotency does not impose dynamical equations on the background metric. This can be cured in the $\mathcal{N}=4$ generalization of the model along the lines of~\cite{Bonezzi:2018box}.

Whether a dynamical system obtained in this way admits an action is a non-trivial question. In order to construct an action one needs to endow the operator algebra $\algA$ with an (invariant) trace with negative degree, which does not naturally arise for this type of algebras. Under certain conditions, which usually amount to an operator-state correspondence between elements of $\algA$ and states of some module of $\algA$, this trace can be realised as an inner product on the latter but turns out to be degenerate. We use this trace to construct a linearized action for which the linearized nilpotency derives from a  variational principle and then discuss possible non-linear generalisations.

There is an alternative approach to the quantization of the relativistic particle, which starts with Brink-Schwarz superparticle \cite{Brink:1981nb}, where space-time- rather than world-line supersymmetry is manifest. Quantization of the latter \cite{Sorokin:1988nj}, after a suitable gauge fixing, gives rise to a BRST operator $\tilde \Omega$ which is a derivation of the standard product of graded functions on the representations space \cite{Berkovits:2001rb}. The latter property allows to construct gauge-invariant polynomial actions on the space of functions with an action of $\tilde\Omega$. In particular, non-abelian super-Yang-Mills theory in 10 dimensions arises as a cubic BV-action in this way \cite{Berkovits:2001rb,Berkovits:2002zk}.

We begin, in section \ref{Lie}, by reviewing how a general graded Lie super algebra gives rise to a BV-BRST gauge system. In section \ref{pointp} we focus on graded Lie algebras arising from the world-line with ($\mathcal{N}$-extended) world-line supersymmetry. However, these algebras do not generically determine on-shell fields. We then describe the reduction of $\algA$, first in general and then, in section \ref{n=1}, applied to the world line with $\mathcal{N}=2$ and $\mathcal{N}=4$ and give a complete description of the theory of background fields for $\mathcal{N}=2$ at filtration zero, thus involving background fields of spin no larger than  one. There we also restore background independence in filtration 1 by considering an arbitrary (off~shell) gravity background and comment on how the latter goes on~shell for $\mathcal{N}=4$. In section \ref{osc} we employ  the operator state correspondence of \cite{Dai:2008bh} to construct a degenerate trace on $\algA$ which is invariant at the linearized level. Nevertheless this trace gives rise to a correct linearized action on $\algA$. We then propose an ad-hoc non-linear extension of which reproduces Yang-Mills theory up to higher derivative terms.

\section{Gauge theories associated to graded Lie ($L_{\infty}$-) algebras}\label{Lie}

For a given graded Lie super algebra $\algA$ with a $\fZ$-grading, called ghost degree, 
there is a natural gauge system associated to $\algA$. To see this, consider a generic element $\Phi$ of degree $1$, Grassmann parity 1, and subject it to the master equation
\begin{equation}
\label{phi-eom}
    \commut{\Phi}{\Phi}=0\,.
\end{equation}
In case there is a non-degenerate invariant inner product of ghost degree $-3$ on $\algA$ this equation follows from a Lagrangian 
\begin{equation}
\label{phi-act}
    L=\frac{1}{6}\inner{\Phi}{[\Phi,\Phi]}\,,
\end{equation}
suggested originally in  \cite{Horowitz:1986dta}. We should emphasise that $\inner{\cdot}{\cdot}$ does not directly  originate form the standard trace since it carries ghost degree $-3$.

In order to describe the gauge symmetries of this system we note that if $\Xi$ is an element of ghost degree $0$ it determines an infinitesimal gauge transformation
\begin{equation}
\label{phi-gs}
    \delta \Phi=\commut{\Phi}{\Xi}\,.
\end{equation}
In particular, it maps solutions of (\ref{phi-eom}) to solutions. Finite transformations are obtained by exponentiation. In a similar way one defines gauge for gauge symmetries. We refer to this gauge theory as the theory associated to $\algA$. 

To continue we wish to associate a BV-BRST homological vector field to this gauge system. 
For the moment we will be interested in the description at the level of equations of motion and their gauge symmetries  so that we disregard the BV master action and the odd symplectic structure, focusing on the homological vector field known as BRST differential.\footnote{We will reconsider the existence of a Lagrangian in section \ref{osc}.} More details on the BV formulation at the level of equations of motion can be found in~\cite{Barnich:2004cr,Kazinski:2005eb}.
In order to construct BV-BRST description to each element $e_A$ of the homogeneous basis in $\algA$ one associates a variable (coordinate) $\psi^A$ of ghost degree $\gh{\psi^A}=-\gh{e_A}+1$ and Grassmann parity $\p{\psi^A}=\p{e_A}+1\,\, mod \,\,2$. These can be interpreted as follows: a generic element from $\algA$ has the form $\phi^A e_A$, where $\phi^A$ are real numbers. Then one promotes each $\phi^A$ to a coordinate $\psi^A$ on a graded supermanifold  $M_\algA$, associated to $\algA$ seen as a graded superspace. The algebra of functions on $M_\algA$ is then a graded super-commutative algebra generated by $\psi^A$.

$M_\algA$ is naturally a $Q$-manifold, with the $Q$-structure determined by the algebraic  structure on $\algA$. To describe the latter, it is useful to define the following canonical linear function on $M_\algA$ with values in $\algA$:
\begin{equation}
 \Psi=\psi^A e_A, \qquad \gh{\Psi}=1, \qquad \p{\Psi}=1\,,
\end{equation} 
which we will call a string field in what follows. The $Q$-structure is then a vector field  $Q=Q^A(\psi)\Dl{\psi^A}$ on $M_\algA$, defined as follows: the action of $Q$ on linear functions is defined through
\begin{equation}
\label{Q-def}
 Q\Psi=\half\commut{\Psi}{\Psi}
\end{equation} 
and then it is extended to all functions by requiring graded Leibnitz rule. $Q$ is nilpotent by construction\footnote{We should stress that $Q$ is not to be confused with the world-line BRST charge whose nilpotence typically implies some equations of motion as we will see in the following sections.}  and the $Q$-manifold $(M_\algA,Q)$ is a graded version of the usual Chevalley-Eilenberg (CE) complex which reduces to the latter if $\algA$ is a usual Lie algebra. In terms of components one has
\begin{equation}
    Q\psi^A=\half U_{BC}^A \psi^B\psi^C\, \quad\text{with}\quad \commut{e_A}{e_B}=U_{AB}^C e_C\,.
\end{equation}
This is the BV-BRST formulation of the background independent  gauge system determined by a graded Lie algebra ${\cal{A}}$. It is background independent in the sense that the system is defined without a reference to any fixed vacuum solution. Moreover, the equations of motion and gauge symmetries are at least quadratic in fields and hence do not arise as a perturbative expansion of a more general system.

If $\algA$ can be represented as an algebra of functions on space-time, with coordinates $x^\mu$, with values in some ``internal'' linear space, it is convenient to introduce a field $\psi^A(x)$ to each basis element $e_A$ of the internal space so that the string field really becomes a field defined on space-time. In this case the Lie bracket is in general a bidifferential operator. $Q$ is then extended to all local functions by requiring $\commut{D_\mu}{Q}=0$, where $D_\mu$ is a total derivative. 

The graded supermanifold $M_\algA$
equipped with $Q$ gives a BV formulation
of the above gauge system.  In particular, equations of motion~\eqref{phi-eom} and gauge symmetries~\eqref{phi-gs} are determined by $Q$.  Indeed, the equations of motion encoded in $Q$ can be obtained as 
$\left(Q\Psi^{(-1)}\right)|_{\Psi^{(k)}=0}=0$, $k\neq 0$, where  $\Psi^{(k)}$ denote a component of a string field containing ghost-degree $k$-fields. Explicitly one gets ($k\neq 0$)
\begin{equation}\label{zeroq}
Q\Psi^{(-1)}|_{\Psi^{(k)}=0}=\half{[\Psi,\Psi]}|_{\Psi^{(k)}=0}=\half\commut{\Psi^{(0)}}{\Psi^{(0)}}\,,
\end{equation}
giving the original equations~\eqref{phi-eom} upon identifying $\Psi^{(0)}$ and $\Phi$.\footnote{This requires some care in the case if fermions are present among ghost degree zero fields.} Similarly, one finds that $Q$ encodes the gauge symmetries~\eqref{phi-gs}. Further details can be found in~\cite{Grigoriev:2006tt}.

Although we won't have a need for it in this paper, in closing this section, let us comment on how the above construction can be generalized to the case where $\algA$ is replaced by an $L_\infty$-algebra ( it goes without saying that we are dealing with $\fZ$-graded superalgebras). Suppose we are given an $L_{\infty}$-algebra determined by the following polylinear maps on the space $\algA$:
\begin{equation}
 \cD:\algA\to\algA\,, \qquad [\cdot,\cdot]:\algA\tensor\algA\to \algA\,, \qquad [\cdot,\cdot,\cdot]:\algA\tensor\algA\tensor\algA \to \algA\,,\qquad \ldots\,,
\end{equation} 
which are super-skew-symmetric, and satisfy $L_\infty$ conditions (i.e. $\cD$ is nilpotent, $[\cdot,\cdot]$  satisfies Jacobi identity  modulo $\cD$-exact terms, etc.). In addition the ghost degree of a $k$-linear map is assumed to be $2-k$. We note, in passing, that the construction also applies to $A_\infty$-algebras just because each $A_\infty$ is an $L_\infty$ as well.

All these structures can also be encoded in the $Q$-structure on $M_\algA$ determined by\footnote{In various contexts this construction appeared in \cite{Zwiebach:1993ie,Alexandrov:1995kv,Grigoriev:2006tt,Erler:2013xta} for instance. }
\begin{equation} \label{Q-general}
Q\Psi=\cD \Psi+\half[\Psi,\Psi]+\frac{1}{6}[\Psi,\Psi,\Psi]+\ldots\,.
\end{equation} 
This defines the action of $Q$ on coordinates $\psi^A$ and hence on generic functions on $M_\algA$.  The nilpotency of $Q$ encodes the $L_\infty$-conditions.
Indeed, acting with $Q$ on both sides of \eqref{Q-general},  one gets $\cD^2\Psi=0$ at linear order in $\Psi$, $\commut{\cD\Psi}{\Psi}-\commut{\Psi}{\cD\Psi}-\cD\commut{\Psi}{\Psi}=0$ at quadratic and so on. To rewrite these conditions in the standard form, where the entries are $\algA$-elements, e.g. basis elements in $\algA$,  rather than string field $\Psi$, one can consider their Taylor coefficients.

With our conventions $Q$ (as a vector field on $M_\algA$) also carries ghost degree $1$. This $Q$ again determines a gauge theory (see e.g.~\cite{Grigoriev:2006tt}). In particular, the explicit form of equations of motion can be obtained from $Q$ as $0=Q\Psi^{(-1)}|_{\Psi^{(l)}=0}$, $l\neq 0$. Explicitly this gives with $\Psi^{(0)}=\Phi$,
\begin{equation}
\label{L-eoms}
    \cD \Phi+\half\commut{\Phi}{\Phi}+\frac{1}{6}[{\Phi},{\Phi},{\Phi}]+\ldots=0\,,
\end{equation}
where the brackets are those determining $L_\infty$-structure. Furthermore, gauge symmetries are determined by $Q$ through
\begin{equation}
\delta_\Xi \Phi=(Q\Psi^{(0)})|_{\Psi^{0}=\Phi,\Psi^{(1)}=\Xi, \Psi^{(k)}=0 \, \, k\neq 0,1\,,}\;,\quad 
\end{equation}
where $\Xi$ is a generic element of ghost degree $0$ (gauge  parameter). Explicitly, this gives
\begin{equation}
    \delta_\Xi \Phi=\cD\Xi+\commut{\Phi}{\Xi}+\frac{1}{2}[{\Phi},{\Phi},{\Xi}]+\ldots\,.
\end{equation}

Conversely, a perturbative gauge theory gives rise to an $L_\infty$-algebra. Indeed, for a perturbative gauge theory the BV supermanifold of fields, antifields and ghosts can be taken to be a linear supermanifold and the theory is determined by a BRST differential $Q$ which is a formal series in coordinates. This data is equivalent to that of an $L_\infty$ algebra through~\eqref{Q-general}.
More abstractly, this can be expressed as a known statement that $L_\infty$ algebras are in 1:1 with formal $Q$-manifolds; the respective  categories are equivalent. 
Although for perturbative gauge theories the $L_\infty$ description can be inferred from the BV one, the $L_\infty$-description can be quite useful (see e.g.~\cite{Jurco:2018sby} and references therein). Let us however mention a few important subtleties:
\begin{itemize}
    
    \item  The above discussion is given at the level of equations of motion. To work with Lagrangian theories
one needs a compatible symplectic structure on the Q-manifold or a suitable invariant inner product on the $L_\infty$-algebra.

\item  The BV formulation is more general because it is well-defined not just for formal manifolds (i.e. perturative theories) and hence is especially relevant in the case of background-independent theories. More precisely, given a $Q$-manifold and a point of the zero locus of $Q$ (vacuum solution), the tangent space at any point of its zero locus is naturally equipped with $L_\infty$ structure~\cite{Alexandrov:1995kv}.  In more physical terms this corresponds to perturbative expansion of the theory around a vacuum solution (the point of zero locus). 
\item  Another subtlety is that if we are dealing with a local gauge field theory its structure is not captured by just $L_\infty$. From the BV perspective the relevant setting is given by the local BV formalism, defined on suitable jet-bundles (see e.g.~\cite{Barnich:2000zw} and references therein). A more invariant setup for local gauge theories at the level of equations of motion is provided by so-called gauge PDEs (also known as parent formulation)~\cite{Barnich:2010sw,Grigoriev:2012xg,Grigoriev:2019ojp}. 
\end{itemize}

\section{The BRST quantized relativistic particle and background fields} \label{pointp}

In the present exploration we will focus on the algebra  $\algA$  that arises from the BRST quantization of the relativistic particle, where the degree is given by the BRST ghost number. Considering $\algA$ as a graded Lie superalgebra and applying the above construction one ends up with the gauge theory with BRST differential $Q$, which defines a theory of background fields on which the particle described by the constraint system could propagate consistently. This is tantamount to nilpotency of the world line BRST operator $\Omega$.

\subsection{Scalar particle}
Let us illustrate this with the example of scalar point particle with no world line SUSY. For this we consider the associative algebra $\algA$ of quantum operators (quantized phase space) determined by
\begin{equation}
 \commut{x^a}{p_b}=i\delta^a_b\,, \qquad  \commut{c}{b}=1\,,
\end{equation} 
where $\commut{\,}{}$ denotes the graded commutator and the Grassmann degree and the ghost degree are assigned as 
\begin{equation}
\gh{x}=\gh{p}=0\,, \qquad 
\gh{c}=-\gh{b}=1\,, 
\end{equation} 
\begin{equation}
\p{x}=\p{p}=0\,, \qquad 
\p{c}=\p{b}=1\,.
\end{equation} 

For a particle propagating on flat Minkowski space-time with metric $\eta_{ab}$ and no  other background fields present the world line BRST operator is given by 
\begin{equation}\label{qb1}
 \Omega^{init} =-c \eta^{ab}p_a p_b\,, \quad \p{\Omega^{init}}=1
\end{equation} 
The general deformation of $\Omega$ can be parametrized as
\begin{equation}\label{Oext0}
 \Omega^{ext}=c (g^{a_1\cdots a_k}p_{a_1}\cdots p_{a_k}+D(x))\,,\qquad k\geq 1\,,
\end{equation} 
and describes conformal (higher-spin) backgrounds (see  \cite{Segal:2002gd,Grigoriev:2006tt}). Gauge transformations (\ref{phi-gs}) amount to a Weyl rescaling of $g_{a_1\cdots a_k}$ in conjunction with a shift in $D$ plus higher spin  diffeomorphisms. Nilpotency of $\Omega$, or equivalently the master equation (\ref{phi-eom}) is trivially satisfied due to $c^2=0$, so that the associated conformal higher spin gauge theory is off-shell.\footnote{This off-shell higher spin theory has been proposed in~\cite{Segal:2002gd} without employing BRST technique. The above BRST description was given  in~\cite{Grigoriev:2006tt}.}

In order to disentangle background fields of different spins we introduce an additional degree for $\algA$, given by 
~\cite{Grigoriev:2012xg} 
\begin{equation}
\begin{gathered}
  \deg{p_a}=1\,,\quad \deg{x^a}=0\,,\quad    \deg{c}=-1\,,\quad \deg{b}=2\,. 
\end{gathered}
\end{equation} 
A reader familiar with string theory will recognize the correspondence with the conformal weights in the world sheet theory. The associated filtration is given by $\ldots\algA_0\subset\algA_1\subset \ldots\subset \algA$, where we denote by $\algA_r\subset \algA$ a subspace of elements of degree $r$ or less
\begin{equation}
 {\algA_r}{\algA_s}\subset \algA_{r+s}\,,\qquad \commut{\algA_r}{\algA_s}\subset \algA_{r+s-1}
\end{equation} 
In particular, $\algA_r$ is a subalgebrab in $\algA$ for $r\leq 0$ and $\algA_r$ is a Lie subalgebra in $\algA$ for $r\leq 1$. An element $F \in \algA_r$ of definite ghost number parametrizes only a finite number of fields. In particular, $\Omega \in \algA_1$ describes a  finite number of low spin fields. 
For the bosonic world line we then have the result that the BRST operator cohomology classes, i.e. non-exact deformations described by $\Omega^{ext}$, are non-empty for all degrees $\geq-1$.

\subsection{Spinning particles}
A way to generate extensions of the above algebra $\algA$ is to consider models with world-line supersymmetry. 
For $\mathcal{N}=1$, $\algA$ is the associative algebra generated by the operators arising from the quantisation of the world line theory which now includes the world line fermions $\theta^\mu$ together with the super reparametrisation ghosts $c,b,\gamma,\beta$, subject to the commutation relations 
\begin{equation}
 \commut{\theta^a}{\theta^b}=2\eta^{ab}\,, \quad
 \commut{\gamma}{\beta}=1 \,, 
\end{equation} 
where $\commut{}{}$ denotes the graded commutator with Grassmann parity 
\begin{equation}\label{Gd}
\p{\gamma}=\p{\beta}=0\,, \qquad 
\p{\theta}=1\,.
\end{equation} 
The ghost degree is 
\begin{equation}\label{gd}
\gh{\theta}=0\,, \quad 
\gh{\gamma}=1\,, \quad \gh{\beta}=-1\,,
\end{equation} 
while the additional degree determining the filtration is assigned according to
\begin{equation}
 \quad \deg{\theta}=\half\,
  \quad \deg{\gamma}=-\half \,,\quad \deg{\beta}=\frac{3}{2} \,.
\end{equation}

It has been known for a long time that, like for the bosonic world line (\ref{qb1}), the constraints implied by (\ref{phi-eom}) do not restrict the gravitational background \cite{Howe:1988ft}. To continue we choose flat Minkowski space. Then $\Omega^{init}$ in \eqref{qb1} is replaced by 
\begin{align}
\label{N1init}
    \Omega^{init}=-cp^2+\gamma\theta^a p_a +b\gamma^2\,.
\end{align}
In filtration zero, the spin 1, taking into account ghost number, Grassmann parity, filtration, the most general Ansatz for $\Omega^{ext}$ is given by 
\begin{equation}\label{Oext1}
 \Omega^{ext}=c (B^a(x)p_a+{F}^{ab}(x)\theta_a\theta_b+D(x)+\E(x)\beta\gamma)+\gamma{A}^a(x) \theta_a\,.
\end{equation} 
Nilpotency of $\Omega$,
\begin{align}
    \Omega^2 = &\;c\gamma\theta^c\left(\left(4F_{[ca]} + i\left(2\partial_aA_c + \partial_cB_a\right)\right)p^a - 4F_{[ac]}A^a + i\partial_cD + \partial^2A_c - iB^a\partial_aA_c\right) \nonumber\\
    & + ic\gamma\partial_{[c}F_{ab]}\theta^c\theta^a\theta^b - \frac{4}{3}ic\gamma\partial^aF_{ca}\theta^c + \frac{2}{3}ic\gamma\partial^aF_{ac}\theta^c + \frac{2}{3}ic\gamma\partial_aF^{\ b}_b\theta^a \nonumber \\
    & +\gamma^2\left(\left(2A^a+B^a\right)p_a - i\partial_aA^a + A^2 + \left(F_{ab}-i\partial_{[a}A_{b]}\right)\theta^a\theta^b + D\right) \nonumber \\
    & -c\gamma\theta^a\E\left(p_a + A_a\right) + ic\beta\gamma^2\theta^a\partial_a\E + ic\gamma\theta^a\partial_a\E + 2bc\E\gamma^2 + \E\beta\gamma^3 = 0,
\end{align}
then implies that $\E=0$, $B_a=-2A_a$, $D=i\partial_a A^a-A^2$ and $F_{ab}=\frac{i}{2}(\partial_a A_b-\partial_b A_a)$ but leaves $A_a$ unconstraint. Thus, as for $\mathcal{N}=0$, gravity and spin one background field are off-shell. More generally, if we do not restrict to particular filtration, generic deformations of \eqref{N1init} are known to describe off-shell  conformal higher spin theory of Type-B~\cite{Grigoriev:2018wrx}.


For $\mathcal{N}=2$, $\algA$ is the associative algebra generated by the operators arising from the quantisation of the world line theory which now includes the world line fermions $\theta^\mu=\frac{1}{\sqrt{2}}(\theta_1^\mu+i \theta_2^\mu)$, $\bar \theta^\mu=\frac{1}{\sqrt{2}}(\theta_1^\mu-i \theta_2^\mu)$ together with the super reparametrisation ghosts $c,b,\gamma,\bar\gamma,\beta,\bar\beta$, subject to the commutation relations 
\begin{equation}
 \commut{\theta^a}{\bar \theta_b}=\delta^a_b\,, \quad
 \commut{\gamma}{\bar\beta}=1 \,, \quad \commut{\bar\gamma}{\beta}=1
\end{equation} 
where $\commut{}{}$ denotes the graded commutator. Grassmann partity and ghost number are the same as given in \eqref{Gd} and \eqref{gd} for $\mathcal{N}=1$. 
  The additional degree determining the filtration is assigned according to
\begin{equation}
 \quad \deg{\theta}=\deg{\bar\theta}=\half\,
  \quad \deg{\gamma}=\deg{\bar\gamma}=-\half \,,\quad \deg{\beta}=\deg{\bar\beta}=\frac{3}{2} \,.
\end{equation}

The constraints implied by (\ref{phi-eom}) do not restrict the gravitational background beyond the condition that the connection is torsion free \cite{Howe:1988ft}. Thus, as for $\mathcal{N}=0$,  the gravitational background remains un-constraint for $\mathcal{N}=2$.  Furthermore, as we will see in section \ref{osg} symmetries of $\Omega^{init}$ include conformal Killings in $4$ space-time dimensions. 

Let us now turn to filtration zero. In filtration zero, in contrast to $\mathcal{N}=1$, the spin 1 background fields are set to zero by nilpotency \cite{Dai:2008bh}. Instead, for $\mathcal{N}=2$ there is a new off-shell scalar background which enters in the form of an anti- hermitean vector potential that is pure gauge, and which can be identified with the dilaton. To see this, let us recall the structure of $\Omega$ 
\begin{align}\label{Omega_gravity}
	\Omega = c\mathcal{D}   +\bar\gamma q +\gamma\bar q  + \gamma\bar\gamma b\,.
\end{align}
We then make the following Ansatz for the supercharges:
\begin{equation}\label{qbarqwithPhi}
q:=\theta^\mu(p_\mu-i\partial_\mu\varphi)\;,\quad \bar q:=\bar\theta^\mu(p_\mu+i\partial_\mu\varphi)\,.
\end{equation}
The algebra of the supercharges then takes the form 
\begin{equation}
\{q,\bar q\}=p^2-\,\partial^2\varphi+\,(\partial\varphi)^2  +2\theta^\mu\bar\theta^{\nu}\,\partial_\mu\partial_\nu\varphi\;,
\end{equation}
This suggests to define the complete BRST operator as \begin{align}\label{QwithPhi}
\Omega&=-c\,(p^2-\,\partial^2\varphi+\,(\partial\varphi)^2  +2\theta^\mu\bar\theta^{\nu}\,\partial_\mu\partial_\nu\varphi)+\gamma\bar q+\bar\gamma q+\bar\gamma\gamma\,b\,\nonumber\\
&\equiv c\mathcal{D}\;+\;\mathcal{G}\;+\;\bar\gamma\gamma\,b\,.
\end{align}
Then $\Omega^2=0$ implies the two equations 
\begin{align}\label{o2z}
&\mathcal{G}^2+\bar\gamma\gamma\,\mathcal{D} =0 \nonumber\\
&[\mathcal{D},\mathcal{G}]=0\,.
\end{align}
The vanishing of the first equation is guaranteed by the choice of non-minimal coupling in $\mathcal{D}$. A short calculation then shows that the second equation is also satisfied  without imposing any condition on $\varphi$. This calculation also shows that for $\mathcal{N}=2$, the filtration -1 background field $D$, which was un-constraint for $\mathcal{N}=0$, is determined by an un-constraint filtration 0 dilaton $\varphi$. 

For $\mathcal{N}=4$, the world line fermions and super ghost are doubled with commutation relations
\begin{equation}
\commut{\theta^\mu_i}{\bar \theta^j_{\nu}}=\delta^j_i\delta^\mu_\nu\,, \qquad  \commut{\gamma_i}{\bar\beta^j}=\delta^j_i \,, \qquad \commut{\bar\gamma^i}{\beta_j}=\delta^i_j\,.
\end{equation} 
It has been known for many years that the $\mathcal{N}=4$ spinning particle cannot be coupled to gravity. In addition the analysis in \cite{Bonezzi:2020jjq} shows that a non-constant dilaton is incompatible with nilpotency of $\Omega$ as well. Thus, although we don't know the operator BRST at filtration $0$ and $1$, we do know that gravity and the dilaton is not contained in it.

\subsection{Constraint reduction of  operator algebras}\label{redA}
What we learned from the world line models with extended supersymmetry, discussed in the last section, is that neither Yang-Mills nor Einstein Gravity feature among the theories of background fields arising from these models. This may seem odd in view of the fact that the world line should, in some sense, be the low energy limit of the string\footnote{The algebra $\algA$ of the spinning particle with $\mathcal{N}=2$ and $\mathcal{N}=4$ world-line SUSY is isomorphic to the zero modes of a twisted world sheet of the superstring  \cite{Bonezzi:2018box}.} from which world-sheet conformal invariance does imply Yang-Mills - and Einstein equations. The solution to this puzzle lies in the observation that the super Lie algebras that arise from spinning world lines admits a  consistent reduction. To see this we may restrict to the case where $\algA$ is an associative algebra, since this is the case for all $\algA$ that we have encountered so far. 

Suppose now $\algA$ comes together with its module $\cH$ and suppose we are interested in a subspace $\cH_0\subset \cH$. Then one can consider a reduced algebra which is constructed in two steps. First one considers a subalgebra $\algA^{\prime\prime}$ of elements preserving $\cH_0$
\begin{equation}
\algA^{\prime\prime}=\{a\in\algA : a\cH_0\subset \cH_0\}
\end{equation}
$\algA^{\prime\prime}$ is clearly a subalgebra in $\algA$.

In the next step one considers a quotient by the ideal in $\algA^{\prime\prime}$ that act trivially on $\cH_0$:
\begin{equation}
\cI=\{a\in\algA^{\prime\prime} : a\cH_0=0\}
\end{equation} 
It is clear that $\cI$ is a left-right ideal in $\algA^{\prime\prime}$ and hence $\algA^{\prime}=\algA^{\prime\prime}/\cI$ is an associative algebra. As we are going to see, starting from graded Heisenberg algebra and its standard module one can build a variety of associative algebras whose associated gauge field theories are known field theories such as Yang-Mills or gravity.

Under suitable assumptions the above reduction can be performed without making use of representation. Indeed, suppose that we are given with certain elements $T_i\in \algA$ such that $\commut{T_i}{T_j}=C_{ij}^k T_k$, i.e. they are in involution. Then one can define
\begin{equation}
\begin{aligned}
\label{q-red-def}
\algA^{\prime\prime}&=\{a\in\algA : \commut{a}{T_i}=A_i^j T_j \}\\
\cI&=\{a\in\algA^{\prime\prime} : a=a^i T_i\}
\end{aligned} 
\end{equation}
This definition of $\algA^{\prime\prime}$ and $\cI$ is equivalent provided  $T_i$ are chosen such that $T_i\phi=0$ single out $\cH_0\subset \cH$ and 
satisfy certain regularity assumptions. This reduction is a version of the quantum Hamiltonian reduction. In particular, the definition~\eqref{q-red-def} is used to define higher-spin algebras~\cite{Eastwood:2002su,Vasiliev:2003ev}.

\section{Field theory of the reduced $\mathcal{N}=2$ spinning particle }\label{n=1}

Here we apply the constraint reduction described above to the case where $\algA$ is an algebra of quantum operators of the  BRST-quantized particle with $\mathcal{N}=2$ supersymmetry and work out explicitly the equations on the background field imposed by nilpotency of $\Omega$. The module $\cH$ is given by wave functions $\phi(x|\theta,c,\beta,\gamma)$ which can be viewed as functions of $x$ with values in polynomials in $\theta,c,\beta,\gamma$.  The generators of $\algA$ are represented on wave functions in $x,\theta,c,\beta,\gamma$ according to
\begin{equation}
 p_a=-i\Dl{x^a}\,, \qquad \bar\theta_a=\Dl{\theta^a}\,, \qquad b=\Dl{c}, \qquad \bar\gamma=\Dl{\beta},\qquad \bar\beta=-\Dl{\gamma}\,.
\end{equation} 
On $\cH$ the above data determines a linear gauge system describing the totally anti-symmetric gauge fields. The standard procedure to construct it is based on an inner product on the representation space realized as 
\begin{equation}
 \inner{\,}{}=\int d^dx dc\;\inner{\,}{}_0\,,  
\end{equation} 
where $\inner{\,}{}_0$ is the Fock space inner product uniquely determined by identifying $\theta^\dagger=\bar \theta$, $\gamma^\dagger=\bar \gamma$, $\beta^\dagger=-\bar \beta$, so that e.g.
$\inner{\gamma}{\beta}_0=1$. The quadratic action is then written as
\begin{equation}
\label{M-action}
    S[\phi]=\half \inner{\phi}{\Omega^{init}\phi}\,, \qquad \gh{\phi}=0\,.
\end{equation}
The corresponding BV master actions is obtained by simply replacing $\phi$ with the associated string field $\Psi$.\footnote{In the case of $\mathcal{N}>2$ supersymmetric particles the associated linear system describes tensor fields of more general symmetry type, including e.g. linearized graviton for $\mathcal{N}=4$.}

In order to define $\algA^\prime$ we introduce the  $\Omega^{init}$-invariant operator
\begin{equation}
 N=\theta\cdot\Dl{\theta}+\gamma\Dl{\gamma}+\beta\Dl{\beta}\,, \qquad \commut{\Omega^{init}}{N}=0\,
\end{equation} 
 counting the $U(1)$ $R$-charge. The subspace $\cH_0$ is then defined by $N\phi=q\phi$ for some fixed $q$ which we will soon set to one in order to single out an irreducible subsystem of spin-1 gauge field. Gauge parameters $\chi$ satisfy $\gh{\chi}=-1$ and 
$N\chi=q\chi$. If we restrict $\phi$ to $\cH_0$ the restricted model indeed describes spin $1$ field. In particular, \eqref{M-action} is equivalent to Maxwell Lagrangian for a complex spin-1 field.

Given the above $\cH_0\subset \cH$ one defines the reduced operator algebra $\algA^\prime$. The filtration on $\algA$ defines that on $\algA^\prime$ and from now we concentrate on the Lie-subalgebra $\algA'_1$ and its associated gauge field theory.
Let us stress that the construction of $\algA'$ and the filtration therein does not require any background fields to be defined on the space-time and hence the associated field theory is background independent. 

However, as we are going to see, on-shell fields are found in $\algA'_0$, while the metric background enters through  $\Omega^{init}$ and is unconstrained by nilpotence. Later on we discuss how this asymmetry can be cured by increasing the amount of world line supersymmetry. In any case it is convenient to decompose $\Omega$ as 
\begin{equation}\label{inex}
\Omega=\Omega^{init}+\Omega^{ext}\,,\quad \Omega^{init}\in \algA'_1\,,\quad \Omega^{ext}\in \algA^\prime_0\,
\end{equation} 
and analyze the dynamics of the gauge fields parametrerizing $\Omega^{ext}$ keeping $\Omega^{init}$ to be a fixed nilpotent
operator. Nilpotency of $\Omega$ in $\algA^\prime$ is then equivalent to 
\begin{equation}
\label{Omega-eom}
 \Omega^2\psi\equiv\left(\commut{\Omega^{init}}{\Omega^{ext}} + \half \commut{\Omega^{ext}}{\Omega^{ext}}\right)\psi=0\,, \qquad \forall \psi \in \cH_0\,. 
\end{equation} 
At the same time the natural gauge transformations for $\Omega^{ext}$ are described by 
\begin{equation}
\label{ogs}
\delta \Omega^{ext}=\commut{\Omega^{init}}{\Xi}\,, \qquad \Xi\in\algA^\prime_0,\,\,\, \gh{\Xi}=0\,.
\end{equation}

\subsection{Maxwell-dilaton system}
Taking into account ghost number, Grassmann parity, filtration, and $\commut{N}{\Omega^{ext}}=0$ the most general Ansatz for $\Omega^{ext}$ in the sense of equivalence classes in  $\algA'$ is given by 
\begin{equation}\label{Oext1}
 \Omega^{ext}=c (B^a(x)p_a+\mathbf{F}^{ab}(x)\theta_a\bar\theta_b+D(x)+\E(x)(\beta\bar\gamma +\gamma\bar\beta))+\gamma(\mathbf{A}^a(x) \bar\theta_a)+ \bar\gamma(\bar{\mathbf{A}}^{a}(x)\theta_a)\,.
\end{equation} 
Here, and in what follows  operators in $\algA'$ are understood as representatives of equivalence classes and equalities are understood as equalities in $\algA'$. In particular, in~\eqref{Oext1} we have suppressed terms proportional to $\beta\bar\gamma-\gamma\bar\beta$ as they  can be reabsorbed into the redefinition of the trace of $\mathbf{F}^{ab}$ by the equivalence relation defining $\algA'$.

As we already discussed the metric enter through $\Omega^{init}$ and we may set it to any background value.  For instance, $\Omega^{init}$ associated to flat Minkowski space is given by
\begin{equation}
 \Omega^{init}
 =-c p^2+\gamma p\cdot\bar\theta+\bar\gamma p \cdot \theta+\gamma\bar\gamma b\,, \quad\gh{\Omega^{init}}=1\,, \; \p{\Omega^{init}}=1\,, \; (\Omega^{init})^\dagger=\Omega^{init}\,
\end{equation} 
and is represented on $\cH$ by 
\begin{equation}\label{oi}
  \Omega^{init}\phi= \left(c\Box-i\gamma \Dl{x}\cdot\Dl{\theta} -i \theta\cdot\Dl{x}\Dl{\beta} + \gamma\Dl{\beta}\Dl{c}\right)\phi\,.
\end{equation}

Now we turn to the detailed study of the fields encoded in $\Omega^{ext}$ and for simplicity set $\Omega^{init}$ to be given by ~\eqref{oi}. More general background metrics will be discussed in Section~\bref{osg}. A tedious, but mechanical calculation shows that with this choice of $\Omega^{init}$ (\ref{Omega-eom}) implies that 
\begin{equation}
\label{ad1}
\begin{aligned}
0&= c\bar{\gamma}\theta^b\left(2i\partial_a\bar{\mathbf{A}}_b + i\partial_bB_a+\mathbf{F}_{ba} - \E\eta_{ab}\right)p^a\\
&+ c\gamma\bar{\theta}^b\left(2i\partial_a\mathbf{A}_b + i\partial_bB_a - \mathbf{F}_{ab} - \E\eta_{ab}\right)p^a \\
& +\gamma\bar{\gamma}\left(B^a + \mathbf{A}^a + \bar{\mathbf{A}}^{a}\right)p_a 
\\
&- 2\gamma\bar{\gamma} cb\;\E
\\
&+ \gamma\bar{\gamma}\left(D - i\partial_a\bar{\mathbf{A}}^{ a} + \bar{\mathbf{A}}_a{\mathbf{A}}^a+ \E \right)
\\
& + c\gamma\bar{\theta}^b\left(i\partial^a\mathbf{F}_{ab} + \partial^a\partial_a\mathbf{A}_b + i\partial_bD-iB^a\partial_a\mathbf{A}_b - \mathbf{F}_{ab}\mathbf{A}^a - {\E}\mathbf{A}_b\right) \\
& + c\bar{\gamma}\theta^a\left(\partial^b\partial_b\bar{\mathbf{A}}_a + i\partial_a\left(D + \E\right)-iB^b\partial_b\bar{\mathbf{A}}_a + \mathbf{F}_{ab}\bar{\mathbf{A}}^{ b} - \E\bar{\mathbf{A}}_a\right)
\end{aligned}
\end{equation}
as an equivalence class in $\algA_1^\prime$.  This gives rise to algebraic relations between the fields as well as equations of motion. Note that each line in (\ref{ad1}) leads to one independent constraint. The first four conditions then imply 
\begin{align}\label{fetc}
&\mathbf{F}_{ab}=i\left(2\d_a \mathbf{A}_b-\d_b \mathbf{A}_a-\d_b \bar{\mathbf{A}}_a \right) \ , \quad \mathbf{F}_{ab}=i\left(-2\d_b \bar{\mathbf{A}}_a+\d_a \mathbf{A}_b+\d_a \bar{\mathbf{A}}_b \right)\ ,\nonumber \\ &B^a=-\mathbf{A}^a-\bar{\mathbf{A}}^{ a} \ , \quad \E=0\ ,
\end{align}
and are equivalent to the condition that $[\Omega^2,f(x,c)]=0$ for arbitrary functions of $x$ and $c$. Note that $\mathbf{F}_{ab}$ has two different expressions. In order for both expressions to be mutually consistent, we find,
\begin{equation}\label{Sta}
    \bar{\mathbf{A}}_a=\mathbf{A}_a + \C_a\,, \qquad \d_a \C_b-\d_b \C_a=0,
\end{equation}
This allows to express $\C_a$ as $-2i\d_a\Phi$ in terms of a potential $\Phi$. 
Using the above relations we find that 
\begin{equation}\label{fmn}
    \mathbf{F}_{ab} = 2i\left(\d_a\mathbf{A}_b-\d_b\mathbf{A}_a\right) -2\d_a\d_b\Phi\,.
\end{equation}
The last three conditions in \eqref{ad1} then fix as $D=i\partial_a\bar {\mathbf{A}}^a-\bar{\mathbf{A}}^a {\mathbf{A}}_a$ and furthermore, imply the equations of motion 
\begin{align}\label{mdi}
    \partial^a(\partial_a \mathbf{A}_b-\partial_b \mathbf{A}_a)=2(\partial_a \mathbf{A}_b-\partial_b \mathbf{A}_a)\partial^a\Phi\,.
\end{align}
Thus, $\Omega^2=0$ encodes the complexified Maxwell equations coupled to an off-shell complex scalar field. The reality condition $\Omega=\Omega^\dagger$ with respect to the natural inner product in $\cH$ implies that $\bar{\mathbf{A}}=\mathbf{A}^\dagger$, that is ${\mathbf{A}}=A+id\varphi$ with $A$ and  $\varphi$ real so that $\varphi$ can be identified with the dilaton in (\ref{qbarqwithPhi}). Then (\ref{mdi}) becomes
\begin{align}
\label{mdir}
    \d_aF^{ab}=2F^{ab}\d_a \varphi\,.
\end{align}
where
\begin{align}
    F_{ab}=\partial_a {A}_b-\partial_b {A}_a\,.
\end{align}
Eqns. \eqref{mdi} and \eqref{mdir} take the form Maxwell's equation in the presence of a background connection parametrized by the dilaton. As a result, \eqref{mdir} is invariant under the gauge  transformation $F^{ab}\to e^{2\lambda}F^{ab}$, $\phi \to \phi+\lambda$ which, however, is not encoded in the natural gauge transformations of $\Omega^{ext}$ as we discuss in section \ref{osg}. 

One may attempt 
to eliminate the dilaton from the spectrum by imposing the reality condition such that    $\mathbf{A}_a,\mathbf{\bar A}_a, B_a$ are real and $\mathbf F,\cE$ imaginary but this is not a canonical choice despite the condition is compatible with~\eqref{ad1}.  A systematic consistent restriction of the system is to impose a reality condition in $\algA'$ and $\cH$ (which becomes pseudo-Euclidean then) and requiring $\Omega$ to preserve the inner product in $\cH$. However, this eliminates the vector potential completely. Finally one may try to embed $\algA'$ in the bigger algebra of the $\mathcal{N}=4$ world line where the dilaton goes on-shell \cite{Bonezzi:2020jjq}.

We already mentioned that natural gauge symmetries of the system are given by~\eqref{ogs}. Taking into account ghost degree and filtration the general form of $\Xi\subset\algA'_0$ is $\Xi = i \lambda$. 
One then obtains,
\begin{align}
    \commut{\Omega^{init}}{\Xi} = c \left(i\partial^a\partial_a \lambda -2\partial_a \lambda p^a\right) + \gamma \partial_a \lambda \bar{\theta}^a + \bar{\gamma} \partial_a \lambda \theta^a\,,
\end{align}
giving
\begin{align}
    \delta A_b = \partial_b \lambda\;,\qquad \delta\varphi=0\,.
\end{align}
We have thus identified an operator algebra $\algA^\prime_1$ such that the most general nilpotent hermitean  $\Omega$ of the structure
$\Omega=\Omega^{init}+\Omega^{ext}$ with $\Omega^{ext}\in \algA^\prime_0 \subset \algA^\prime_1$ describes Maxwell theory coupled to an off-shell dilaton. 

\subsection{Non-abelian extension and Yang-Mills fields}

Here we consider the non-abelian generalization of (\ref{ad1}). To this end, as an underlying associative algebra $\hat \algA$ (replacing $\algA$) we take the tensor product of  $\algA$ 
with a universal enveloping algebra $\cU^\algg$ of a given Lie algebra $\algg$. The algebra $U^\algg$ is equipped with the standard filtration $U^\algg_0\subset U^\algg_1 \subset U^\algg_2\ldots$, where $U^\algg_k$ contains elements of order at most $k$ in the generators $T_a$ (basis elements of $\algg$).  In particular, $U^\algg_0$ is the 1-dimensional subalgebra while $U^\algg_1/U^\algg_0$ seen as a Lie algebra is $\algg$ itself. Furthermore, just like for $\algA$ we have
\begin{align}
 U^\algg_mU^\algg_n\subset U^\algg_{m+n}\,, \qquad 
\commut{U^\algg_m}{U^\algg_n}\subset U^\algg_{m+n-1}\,. 
\end{align}
Now taking as $\hat\algA$ the tensor product of $\algA$ and $\cU^\algg$ we obtain an associative algebra analogous to $\algA$. In particular, $\hat\algA$ is equipped with the total filtration induced from the  filtration on $\algA$ and $\cU^\algg$ and, moreover, $\hat\algA^\prime_1$ is again a Lie subalgebra, giving a consistent gauge theory determined by $\Omega^2=0$ for $\Omega \in\hat\algA^\prime_1$. Again, the construction of $\hat\algA^\prime_1$ does not involve background fields and hence the resulting theory is background-independent.

Now we decompose $\Omega$ with respect to the filtration in $\algA$ factor as
\begin{equation}
\Omega=\Omega^{init}+\Omega^{ext}
\end{equation}
so that $\Omega^{init}$ has $\algA$-filtration $1$ and  $\Omega^{ext}$ has $\algA$-filtration $0$. If in addition $\Omega^{init}$ has
$\algA$-degree $1$ it must be of the form $\Omega^{init}\tensor \mathbf{1}$. So it  automatically happens that $\Omega^{init}$ is Abelian. At the same time, because $\Omega^{ext}$ is of total filtration $1$ and $\algA$-filtration $0$ it can be interpreted as a $\cU^\algg_1$ valued element of $\algA$ and, as such, can be decomposed into an Abelian- (i.e. $\cU^\algg{}_0$-valued) and $\algg$-valued one.

As far as the explicit computations are concerned, the nonabelian generalization amounts to simply taking all the coefficient fields entering $\Omega^{ext}$ to be matrix-valued and $\Omega^{init}$ proportional to the unit matrix. To study the equations of motion encoded in $\commut{\Omega}{\Omega}=0$ we can simply copy from the last subsection the terms which are linear in the background fields since they are unchanged for matrix valued fields. In particular the 3rd and 4th line in (\ref{ad1}) imply that the constraints
\begin{align}\label{bna}
&B^a=-\mathbf{A}^a-\bar{\mathbf{A}}^{a}\,,\quad \E=0\,,
\end{align}
are unchanged. After substitution of (\ref{bna}) the generalization of the first lines in  (\ref{ad1}) for matrix valued background fields is found to be  
\begin{align}
 0&=c\bar\gamma \theta^b([\bar{\mathbf{\Pi}}_b, \mathbf{\Pi}_a+\bar{\mathbf{\Pi}}_a ]+ \mathbf{F}_{ba})p^a\\
 &
 +c\gamma\bar \theta^b([\mathbf{\Pi}_b, \mathbf{\Pi}_a+\bar{\mathbf{\Pi}}_a]- \mathbf{F}_{ab})p^a\,,
 \end{align}
where $\mathbf{\Pi}_a=p_a+\mathbf{A}_a$. 
This implies,
\begin{align}
    2\mathbf{F}_{[a,b]}&=[\mathbf{\Pi}_b,\mathbf{\Pi}_a]+[\bar{\mathbf{\Pi}}_b,\bar{\mathbf{\Pi}}_a]+[\mathbf{\Pi}_b,\bar{\mathbf{\Pi}}_a]+[\bar{\mathbf{\Pi}}_b,\mathbf{\Pi}_a]\nonumber\\
     2\mathbf{F}_{(a,b)}&=[\mathbf{\Pi}_b,\mathbf{\Pi}_a]-[\bar{\mathbf{\Pi}}_b,\bar{\mathbf{\Pi}}_a]+[\mathbf{\Pi}_b,\bar{\mathbf{\Pi}}_a]-[\bar{\mathbf{\Pi}}_b,\mathbf{\Pi}_a]
\end{align}
Symmetry in $(a,b)$ of the second equation implies, in particular, that 
\begin{align}\label{ppd}
    [\mathbf{\Pi}_b,\mathbf{\Pi}_a]=[\bar{\mathbf{\Pi}}_b,\bar{\mathbf{\Pi}}_a]
\end{align}
which is clearly satisfied if $\bar{\mathbf{A}}_a=\mathbf{A}_a$. Assuming this in what follows, the forth line of (\ref{ad1}), which is unchanged, imposes the constraint 
\begin{align}\label{ad4}
 D=  i\partial_a{\mathbf{A}}^{ a} - \mathbf{A}^a{\mathbf{A}}_a \,,
\end{align}
The non-abelian generalization of last two constraints in  (\ref{ad1}) then impose the equations of motion for the remaining independent  variable ${\mathbf{A}}^{ a}$, 
\begin{align}\label{Dai}
    \mathbf{\nabla}^a(\partial_a \mathbf{A}_b-\partial_b \mathbf{A}_a+i[\mathbf{A}_a,\mathbf{A}_b])=0 \,,
\end{align}
so that we recover the Yang-Mills equation of motion as in  \cite{Dai:2008bh}.

It is clear that there are generalizations of the solution just presented. We are not aware of the general solution\footnote{See \cite{Mostow:1980yb} for a thourough discussion of solutions to (\ref{ppd}).} of  (\ref{ppd}) compatible with the remaining constraints. Let us just point that (\ref{ppd}) is solved by 
\begin{align}\label{solm}
    \bar{\mathbf{\Pi}}_a=g^{-1}\mathbf{\Pi}_ag
\end{align}
provided $g$ commutes with $\algg$. With this Ansatz the equation (\ref{Dai}) becomes instead 
\begin{align}
    \mathbf{\nabla}^a(\partial_a \mathbf{A}_b-\partial_b \mathbf{A}_a+i[\mathbf{A}_a,\mathbf{A}_b])=i(\partial_a \mathbf{A}_b-\partial_b \mathbf{A}_a+i[\mathbf{A}_a,\mathbf{A}_b])(\bar{\mathbf{A}}_a-\mathbf{A}_a)\,,
\end{align}
This is a non-abelian generalization of (\ref{mdi}) and describes complexified Yang-Mills theory coupled to a non-abelian, but $\algg$-invariant Dilaton parametrised by $\bar{\mathbf{A}}_a-\mathbf{A}_a$.

\subsection{Off-Shell Gravity}\label{osg}

In this subsection we generalize the derivation in section \ref{n=1} to allow for a generic background metric. For this  take $\Omega_g^{init}\in \algA_1$ to be of the form 
\begin{align} \label{Omega_gravity}
	\Omega_g^{init} = c\mathcal{D} +\bar{\gamma}\theta^\mu	\Pi_\mu+\gamma\bar{\theta}^\mu\Pi_\mu + \gamma\bar\gamma b
\end{align}
where $	\Pi_\mu = p_\mu -i\omega_{\mu ab}\theta^a\bar{\theta}^b \nonumber$ and
\begin{align}
	\mathcal{D} &= \Pi^2 + R_{\mu\nu\lambda\sigma}\theta^\mu\bar{\theta}^\nu\theta^\lambda\bar{\theta}^\sigma \,,\quad
	\Pi^2 =-g^{\mu\nu} \Pi_\mu \Pi_\nu + ig^{\mu\nu}\Gamma^\lambda_{\mu\nu}\Pi_\lambda.\nonumber
\end{align}
Here, $x^\mu$ and $\theta^a$ are regarded as independent variables while $\theta^\mu=e^\mu_a\theta^a$. Note that now we work in general space-time coordinates $x^\mu$ and keep Latin indices to denote the orthonormal basis in the flat tangent space.  One finds \cite{Howe:1988ft} (see also~\cite{Getzler:2016fek} for a recent discussion of the spinning particles on curved backgrounds) that $\Omega^{init}_g$ is nilpotent, provided the connection is torsionless. 

As before, we deform $\Omega$ as 
\begin{align}
	\Omega = \Omega_g^{init} + \Omega^{ext},	
\end{align}
with
\begin{align}
	\Omega^{ext} =& c\left(-2A^\mu\Pi_\mu + e^\mu_a e^\nu_b\left(F_{\mu\nu} - 2D_\mu\partial_\nu\varphi\right)\theta^a\theta^b + iD_\mu A^\mu - A^2 +D^\mu \partial_\mu\varphi - (\partial_\mu \varphi)^2\right) \nonumber\\
	&+ \gamma e^\mu_a \left(A_\mu + i\partial_\mu\varphi\right)\bar{\theta}^a  + \bar{\gamma} e^\mu_a \left(A_\mu - i\partial_\mu\varphi\right) \theta^a,
\end{align}
where $A_\mu$ and $\varphi$ are assumed real and $D_\mu$ is the Levi-Civita covariant derivative acting on tensor fields. From nilpotency of the BRST operator restricted to $\algA'_1$, we find,
\begin{align}
    \Omega^2 = c\bar{\gamma}\theta^\sigma\left(D^\nu F_{\nu\sigma}-2F_{\nu\sigma}\partial^\nu\varphi\right) -  c\gamma\bar{\theta}^\sigma\left(D^\nu F_{\nu\sigma}-2F_{\nu\sigma}\partial^\nu\varphi\right)
\end{align}
From which one can read off the background field equations,
\begin{align}
    D^\nu F_{\nu\sigma}=2F_{\nu\sigma}\partial^\nu\varphi\,,
\end{align}
which we recognise as the Maxwell equations coupled to off-shell metric and Dilaton fields. Note that as expected we do not find the gravitational background field equations. Those do arise when considering the $\mathcal{N} = 4$ relativistic particle \cite{Bonezzi:2018box,Bonezzi:2020jjq}.

It is well known that in four dimensions Maxwell theory is conformally invariant. 
An efficient way to identify global symmetries in the present approach is to look for the "zero modes" of the  gauge transformation (\ref{phi-gs}) that correspond to symmetries of $\Omega^{init}$.
For simplicity, we consider $\Omega^{init}$ with Minkowski metric. After some trial and error we come up with Ansatz
\begin{align}\label{AnG}
    \Xi_c = i{\xi _a }{p^a } +\frac{1}{2}\partial_a\xi^a + {t_{a b}}\left( {{\theta ^a }{\bar{\theta} ^b } - {{ \theta }^b }{{\bar \theta }^a }} \right)  - \lambda \left( {bc - cb} \right) - \hat f\left( {\gamma \bar \beta  + \beta \bar \gamma } \right) + ic{\hat A_a }\left( {\beta {{\bar \theta }^a } + \bar \beta {\theta ^a }} \right)
\end{align}
 for $\Xi_c\in {\cal{A}}_1$. This leads to 
 \begin{align}
    \commut{\Omega^{init}}{\Xi_c} =& - c\left( {2{\partial _b }{\xi _a }{p^a }{p^b } - i{\partial ^e} \left( {\partial _e }{\xi _d } + \partial_d \xi_e\right){p^d }}  - \frac{1}{2}\partial_e \partial^e \partial_d \xi^d\right) \nonumber\\
    &+ {\partial _a}{\xi _d }\left( {\gamma {{\bar \theta }^a} + \bar \gamma {\theta ^a}} \right){p^d } {- \frac{i}{2}\left( \gamma \bar \theta^a + \bar \gamma \theta ^a \right)\partial_a\partial_d\xi^d} \nonumber \\
   & + \left( {c\left( {2i{\partial_a }{t_{d e }}{p^a } + {\partial^a }{\partial_a }{t_{d e }}} \right) - i\left( {\gamma {{\bar \theta }^a} + \bar \gamma {\theta ^a}} \right){\partial _a}{t_{d e }}} \right)\left( {{\theta ^d }{{\bar \theta }^e } - {\theta ^e }{{\bar \theta }^d }} \right) \nonumber \\
    & + 2{t_{ae }}\left( {\gamma {{\bar \theta }^e } + \bar \gamma {\theta ^e }} \right){p^a}\label{ckg}\\
    &+c\left( {2\lambda {p^2} - {\partial^a }{\partial_a }\lambda  - 2i{\partial_a }\lambda {p^a } + i{{\hat A}_d }\left( {{\theta ^d }{{\bar \theta }^a} - {\theta ^a}{{\bar \theta }^d }} \right){p_a} - {\partial _a}{{\hat A}_d }{\theta ^a}{{\bar \theta }^d }} \right) \nonumber \\
    &  - c\gamma \bar \beta {\partial _a}{\hat A^a} - c\left( {{\partial^a }{\partial_a }\hat f + i\left( {2{\partial_a }\hat f + {{\hat A}_a }} \right){p^a }} \right)\left( {\gamma \bar \beta  + \beta \bar \gamma } \right) \nonumber \\
    & + \frac{i}{2}\left( {2{\partial _a}\lambda  + {{\hat A}_a}} \right)\left( {\gamma {{\bar \theta }^a} + \bar \gamma {\theta ^a}} \right)\left( {bc - cb} \right) + 2\gamma \bar \gamma b\left( {\lambda  - \hat f} \right) \nonumber \\
    &{ + \frac{i}{2}\bar \gamma {\theta ^a}\left( {2{\partial_a }\hat f - {{\hat A}_a}} \right) + \frac{i}{2}\gamma {\bar \theta ^d }{\hat A_d }} - \left( {\gamma {{\bar \theta }^a} + \bar \gamma {\theta ^a}} \right)\hat f{p_a}\nonumber
\end{align}
For $\lambda=\hat f=\hat A_a=0$ and $  t_{ab} = \frac{1}{4} \left(\partial_a\xi_b - \partial_b\xi_a\right) $ the commutator $\commut{\Omega^{init}}{\Xi_c}$ then vanishes on the solutions of the  Killing equation $\partial_a \xi_b + \partial_b \xi_a =0$, in agreement with Poincar\'e invariance. For $d = 4$ with 
\begin{align}\label{idc}
    & \lambda = \hat f = \frac{1}{4}\partial_a \xi^a,\qquad \ \hat A_a = -2\partial_a \lambda, \qquad \  t_{ab} = \frac{1}{4} \left(\partial_a\xi_b - \partial_b\xi_a\right) ,
\end{align}
the vanishing of (\ref{ckg}) implies the conformal Killing equation $ \partial_a \xi_b + \partial_b \xi_a = \frac{1}{2}\partial_c \xi^c \eta_{ab}$ in agreement with conformal invariance the Maxwell action in four dimensions.

As an aside we note that if we consider gauge transformations
$\commut{\Omega^{init}}{\Xi_c}$ with the parameter $\Xi_c$
determined through~\eqref{idc} by $\xi^a$ that is not necessarily a conformal Killing, but still satisfying $\Box\partial_a\xi^a = 0$, one finds that the gauge variation of $\Omega$ is equivalent to
\begin{align}
    \delta g_{ab} = - \partial_a \xi_b - \partial_b \xi_a + \frac{1}{4}\partial_c\xi^c\eta_{ab}\,.
\end{align}

If we consider generic $\Xi^a$ the respective gauge transformation do not amount to the transformation of the metric so that further background fields are to be introduce in $\Omega^{init}$ in order to realize such gauge transformations. The explicit identification of a a complete set of background fields remains an open problem.

\subsection {$\mathcal{N}=4$ extension}
As another illustration of the general reduction described at the beginning of this section we revisit the relativistic particle with $\mathcal{N}=4$ world line supersymmetry briefly described in section \ref{pointp}. For $\mathcal{N}=4$ there is more freedom in choosing the subspace $\cH_0$, corresponding to different choices of generators of the $SO(4)$ R-symmetry present in this model. For the maximal reduction we impose that \cite{Bonezzi:2018box}
 
\begin{equation}
\begin{split}
N_i &:= \theta_i\cdot\Dl{\theta_i}+\gamma_i\Dl{\gamma_i}+\beta_i\Dl{\beta_i} \;,\quad i=1,2\\[2mm]
Y &:= \theta^1\cdot\Dl{\theta_2}+\gamma^1\Dl{\gamma_2}+\beta^1\Dl{\beta_2}\;
\qquad\text{and}\\[2mm]
T &:=\frac{\partial^2}{\partial \theta_1\cdot \partial \theta_1 }+\frac{\partial^2}{\partial \gamma_1 \partial \beta_2 }-\frac{\partial^2}{\partial \gamma_2 \partial \beta_2 } \;
\end{split}    
\end{equation}
commute with $\Omega$. Furthermore, $\cH_0\subset \cH$ is defined by  $N_i\phi=\phi$, $i=1,2$ and  $Y\phi=T\phi=0$ for $\phi\in \cH_0$. This defines the reduced operator algebra $\algA^\prime$ along the same lines as in section \ref{redA}. The filtration on $\algA$ is introduced in a similar way, giving us a Lie subalgebra $\algA'_1$ and its associated gauge field theory.
Let us stress that the construction of $\algA'_1$ again does not require specification of background fields and hence the associated field theory is background independent.  While we do not have a full characterization of the spectrum in filtration 1 we know that the metric is constrained by nilpotency of $\Omega \in \algA'_1$ to be Ricci flat \cite{Bonezzi:2018box}.

Another consistent choice of a subspace $\tilde\cH_0\supset \cH_0$ and the corresponding $\algA'$, is defined by $N_i\phi=\phi$, $i=1,2$ and  $Y\phi=0$ \cite{Bonezzi:2018box}. For this choice of $\algA'$, nilpotency of $\Omega$ implies the coupled equations \cite{Bonezzi:2020jjq}
\begin{equation}
    R_{\mu\nu}-\Lambda g_{\mu\nu}+2\nabla_\mu\nabla_\nu\varphi=0\;,\quad \nabla^2\varphi-2\,\nabla^\mu\varphi\,\nabla_\mu\varphi + 2\Lambda \varphi=0 \;, 
 \end{equation}
for the metric in degree 1, the dilaton in degree 0 and the cosmological constant $\Lambda$ in degree -1. Note that the Maxwell field is set to zero by nilpotency for all choices of $\cH_0$. This can be seen by adding a vector potential to $\Omega$ describing a Minkowski background, that is 
\begin{align}\label{paf}
    q_i &= \theta_i^a\left(p_a + A_a \right) \quad\text{and}\quad
    \bar{q}^j = \bar{\theta}^{i \,a}\left(p_a + A_a \right) \,
\end{align}
with commutator,
\begin{align}
    \bar{\gamma}^i\gamma_j\left\{q_i, \bar{q}^j\right\} = \bar{\gamma}^i\gamma_i H +  \frac{i}{2}\bar{\gamma}^i\gamma_j\theta_i^a\bar{\theta}^{j \,b}F_{ab} ,
\end{align}
where $H = p^2 + 2A^a p_a - i\partial_a A^a + A^2$ and $F_{ab} = 2i\left(\partial_a A_b - \partial_b A_a\right)$. Nilpotency of $\Omega$ in $\algA^\prime$ then implies $F_{ab} = 0$.

\section{Operator-state correspondence, physical states and action} \label{osc}

Having obtained the Yang-Mills equations of motion as a consequence of the nilpotency of $\Omega$ we will now address the question of constructing an action from which these equations, or equivalently, \eqref{phi-eom} derive. The field content encoded in $\Omega^{ext}\in \algA'_0$ does not seem to admit an invariant inner product required for the action principle so our strategy is to split  the component equations encoded in ~\eqref{ad1} into off-shell algebraic constraints and the remaining equations. More specifically, the off-shell constraints are encoded in
\begin{equation}
\label{alg-const}
\commut{\Omega^2}{f(x,c)}=0    
\end{equation}
for any $f(x,c)$. These encode\footnote{Strictly speaking they are not algebraic because one of the equations is $d(\bar \bA-\bA)=0$. However, 
in terms of our parameterization $\bar \bA-\bA=d\Phi$ the differential constraint is not present.}
the first four lines of~\eqref{ad1} and are sufficient to eliminate all the fields save for $\mathbf{A},\Phi$ and $D$. 

\subsection{Inner product and quadratic Lagrangians}

We now look for an action functional for $\Omega^{ext}$ such that, upon eliminating all the fields save for $\bA,\Phi,D$  via~\eqref{alg-const}, 
its Euler-Lagrange equations reproduce the remaining equations from~\eqref{ad1}, i.e. Maxwell equations on the dilaton background. For this purpose we need an invariant inner product on $\algA^\prime_0$ which, thanks to the algebraic constraints, could well be degenerate.  More precisely, it should become nondegenerate only after imposing these constraints and disregarding the dilaton $\Phi$ because it is off-shell.

In order to construct such an inner product, following \cite{Dai:2008bh}, we first observe that there exists a map  $\mu$ from $\algA^\prime$ to $\mathcal{H}_0$ that can be used to pull the standard inner product on $\mathcal{H}_0$ back to $\algA^\prime$. To construct this map we pick an element in $\mathcal{H}_0$ represented by the wave function $\beta$ (as before here we use the identification of the Fock space vectors and functions in creation operators $\beta,\gamma,\theta,\bar\theta$).  Thanks to 
$\Omega^{init} \beta=0$ it is in the cohomology of $\Omega^{init}$ of ghost degree -1.  Then for a generic element $O\in \algA''$ the map
$\mu$ is defined as
\begin{equation}
\mu (O)=O\beta    
\end{equation}
and it induces a well-defined map  $\algA'\to \cH_0$ because $\beta\in\cH_0$ so that $\mu$ acts trivially on $\cI \subset \algA''$. Finally the degenerate inner product on $\algA^\prime$
is defined as:
\begin{equation}
\inner{a}{b}_{\boldsymbol{\beta}}:=\inner{a \boldsymbol{\beta}}{ b \boldsymbol{\beta}}\,. 
\end{equation}
It turns out this inner product is nondegenerate on a subspace of  $\algA'_0$ singled out by the constraints~\eqref{alg-const} and with the dilaton set to a fixed background value. Indeed, generic $\Omega^{ext}\in \algA'_0$ that belongs to the subspace is parameterized by the unconstrained complex $\bar\bA_a,D$ and is mapped by $\mu$ to 
\begin{align}\label{lina}
    \psi=\mu(\Omega^{ext})=\Omega^{ext}\beta=\bar\bA_b\theta^b+D c\beta\,,
\end{align}
which is a generic ghost degree $0$ element of $\cH_0$ so the map is indeed 1:1. Analogous considerations apply to other ghost degrees so that the inner product is indeed nondegenerate.

Unfortunately such an inner product is not invariant, i.e. in general for an antihermitean $a$
\begin{equation}\label{in1}
    \inner{\commut{a}{c}}{b}_{\boldsymbol{\beta}}+
    \inner{c}{\commut{a}{b}}_{\boldsymbol{\beta}}\neq 0\,.
\end{equation}
However, if $a \boldsymbol{\beta}=0$ the above becomes an equality. This is the property we employ to construct the linearized action. Indeed, consider the quadratic approximation of $\frac{1}{4}\inner{\Omega}{\commut{\Omega}{\Omega}}_{\bbeta}$, 
\begin{equation}
\label{cube-linear}
S[\Omega^{ext}]=\half\inner{\Omega^{ext}}{\Omega^{init}\Omega^{ext}}_{\bbeta}
\end{equation}
where for the moment we do not assume $\Omega^{ext}$ hermitean.
This action is by construction gauge invariant and using \eqref{lina} and the algebraic constraints can be rewritten as the standard  $\inner{\psi}{\Omega^{init}\psi}$ action (c.f. ~\eqref{M-action}) for the spin-1 field. Explicitly one gets
\begin{align}\label{qm2}
  S[\Omega^{ext}] &=-\frac{1}{2}\int d^dx\;\left(\bar\bA^\dagger_b\Box \bar\bA_b +iD^\dagger (\partial\cdot  \bar\bA) +i\bar\bA^\dagger_b \d^b D-D^\dagger D\right)
\end{align}
which upon elimination of the auxiliaries $D,D^\dagger$ gives the standard complexified Maxwell action.  Note that the action does not depend on the dilaton $\Phi$ and hence imposes no equations on it while the complexified Maxwell equation it determines coincides with the linearization of~\eqref{mdi} as expected. Note that (\ref{qm2}) holds irrespective whether we assume the algebraic constraint (\ref{alg-const}) or not. In particular we can impose (\ref{alg-const}) before or after obtaining the equations of motion from (\ref{qm2}).

As we have seen hermiticity on $\Omega^{ext}$ together with
\eqref{alg-const} implies that $\bar\bA_a=A_a-i\d_a\phi$ with $A_a,\phi$ real so that the above complexified Maxwell action reduces to the one for the real field $A_a$ while the dependence on the dilaton drops out because $\phi$ does not contribute to the Faraday tensor. In particular, the resulting equation of motion do not involve dilaton and is a lineraization of the real Maxwell-dilaton system~\eqref{mdir}.

\subsection{Non-linear extension}
In this last subsection we describe an attempt of a non-linear generalization of (\ref{cube-linear}). Here we will assume $\bar{\mathbf{A}}=\mathbf{A}$ and hermitean so that, in particular, the dilaton is absent by assumption. In addition we assume the algebraic constraint (\ref{alg-const}). Then $\Omega$ takes the simpler form 
\begin{align}
    \Omega=-c(\Pi^2+2[\Pi_a,\Pi_b]\theta^a\bar\theta^b-\tilde D)+\bar\gamma\theta^\mu\Pi_\mu+\gamma\bar\theta^\mu\Pi_\mu+b\bar\gamma\gamma\,,
\end{align}
where $\tilde D=D-i\partial\cdot A+A^2$ is the shifted auxiliary field which we assume complex in what follows. The natural extension\footnote{The pefactor $\frac{1}{4}$ instead of the expected $\frac{1}{6}$ from \eqref{phi-act} originates in the non-invariance of the inner product.} of (\ref{cube-linear}) 
\begin{equation}
\label{cube=n}
S[\Omega]=\frac{1}{4}\inner{\Omega}{\commut{\Omega}{\Omega}}_{\bbeta}
\end{equation}
then evaluates to
\begin{multline}
   \frac{1}{4}\inner{\Omega}{[\Omega,\Omega]}{}_\bbeta=-\frac{1}{2}\int d^dx \;\textbf{tr}_{\mathfrak{u}} \left( (\Pi^2-\tilde D^\dagger) \Pi\cdot\Pi-(\Pi^2-\tilde D^\dagger)(\Pi^2-\tilde D)-\Pi^\mu(\Pi^2-\tilde D)\Pi_\mu\right.\\
    \left.\qquad\qquad\qquad+ \Pi\cdot\Pi(\Pi^2-\tilde D) -2\Pi^\mu[\Pi_\mu,\Pi_\nu]\Pi^\nu\right)
\end{multline}
and, after elimination of $\tilde D$ and $\tilde D^\dagger$, 
\begin{align}
\label{t1}
   S[\Omega] 
   =
   \frac{1}{2}\int d^dx \;\textbf{tr}_{\mathfrak{u}} \left( \Pi^\mu\Pi^2\Pi_\mu-\Pi^2\Pi^2 +2\Pi^\mu[\Pi_\mu,\Pi_\nu]\Pi^\nu\right)\;. 
\end{align}
In the commutative case this reproduces (\ref{qm2}), but for a non-abelian gauge group the relative factors don't match to reproduce the Yang-Mills action. This is a consequence of the lack of cyclicity of (\ref{in1}) that does not allow us to bring the first line in (\ref{t1}) into the form $[\Pi_\mu,\Pi_\nu][\Pi^\mu,\Pi^\nu]$. 
We may try to remedy this by starting with an alternative inner product given by 
\begin{equation}
\inner{a}{b}'{}_{\boldsymbol{\beta}}:=\int d^dx d^dy \inner{x|a \boldsymbol{\beta}}{ b \boldsymbol{\beta}|y}K_\epsilon (x,y)\,,
\end{equation}
where $K_\epsilon (x,y)$ is some regulator function to be determined. For now we just assume that it is compatible with the required  cyclicity so that we can cyclicly permute $\Pi^\mu$ and get 
\begin{align}
    \tr(\Omega^\dagger[\Omega,\Omega])=\frac{2}{N}\int d^dx d^dy\;\textbf{tr}_{\mathfrak{u}} <x|[\Pi^\nu,\Pi^\mu][\Pi_\mu,\Pi_\nu]+\tilde D^\dagger\tilde D- \Pi^2\tilde D |y> K_\epsilon(x,y)
\end{align}
where $N= \int d^dx K_\epsilon(x,0)$ is a normalisation factor.
Variation w.r.t. $\tilde D^\dagger$ then gives $\tilde D=0$ and thus produces the expected Yang-Mills action in the absence of the dilaton. 

This still leaves us with the task of finding a regulator function $K_\epsilon (x,y)$ which commutes with $\Pi^\mu$ to ensure the assumed  cyclicity. A natural Ansatz for $K_\epsilon(x,y)$ that comes to mind is
\begin{align}
    K_\epsilon(x,y)=<x|e^{-\epsilon \Pi^2}|y>\stackrel{\epsilon \to 0}{\to}\delta(x-y)\,.
\end{align}
This regulator does not commute with $\Pi^\mu$, due to 
\begin{align}
    [\Pi^\mu,\Pi^2]=-\frac12\left(\Pi^\nu F_{\mu\nu}+F^\mu_{\;\;\nu}\Pi^\mu\right)
\end{align}
However, the extra terms generated by this are suppressed by as $\epsilon^{r+s}\nabla_\lambda^r (F_{\mu\nu})^s$. Thus, we may say that the trace regulated in this way is cyclic (on $\Pi^\mu$) up to terms of higher orders in derivatives or fields (or both).\footnote{We may note that the Ansatz $K_\epsilon(x,y)=<x|e^{-\epsilon p^2}|y>$ works equally well to leading order in $\epsilon$.}

\section{Conclusions}

The purpose of this work was to propose a precise algebraic framework allowing to interpret the results in ~\cite{Dai:2008bh,Bonezzi:2018box,Bonezzi:2010jr} as a toy model for a background independent formulation of string field theory. More specifically, we employed the BV-BRST description of the background fields associated to quantum constrained systems which allows to associate to a given  graded operator algebra $\algA$ (and more generally $L_\infty$-algebra) a gauge field theory in the BV-BRST formulation.  If $\algA$ is an operator algebra of a quantized (spinning) particle, the associated field theory is usually off-shell in the sense that it only implies algebraic constraints and gauge symmetries but not dynamical equations.

For the spinning particle with $\mathcal{N}=2$ supersymmetry there is a natural constraint reduction of $\algA$ associated to the gauging of the world line $R$-symmetry. The space of background fields associated to this algebra is infinite reflecting, in particular, higher spin backgrounds. We then introduced a suitable filtration on $\algA$ which allows to identify, in an invariant way, a finite subalgebra, whose associated fields to not involve infinite tower of higher spins. We explicitly analyzed gauge fields associated to the filtration 0 subalgebra and find that they are exhausted by the  Maxwell field supplemented by an off-shell dilaton.  We also considered the nonabelian extension of the model leading to a Yang-Mills generalization of the system. We did not work out the complete operator BRST-cohomology for all ghost numbers for this problem, but instead focused on the physical fields. Working out the later would be an interesting extension of the present note.

Increasing the amount of world-line symmetry leads to interesting generalizations of $\algA$. In particular, the $\mathcal{N}=4$ particle is known to give rise to on-shell gravity and its extensions. The complete description of the underlying gauge theory and its manifestly background formulation appears to be possible but industrious. Finding a suitable parametrization for this problem would be another interesting extension of the the present work. We note, in passing, that  background fields associated to extended supersymmetric  particles with generic even $\mathcal{N}$ have been discussed in~\cite{Bekaert:2009fg} from the higher spin theory perspective.

Finally, we explored the possibility of constructing a Lagrangian formulation of the gauge theory associated to 
$\mathcal{N}=2$ particle. Generically, this requires an invarinat inner product of degree -3 on $\algA$ as a new ingredient to the formalism. In the quadratic approximation this turns out to be rather straightforward and instructive. On the other hand, we have not succeeded in constructing a satisfactory description of the latter at the non-linear level due to the lack of a cyclic property of our construction. This is in contrast to string theory or the pure spinor approach, where a cubic action on the representation space is known \cite{Berkovits:2001rb,Berkovits:2003ny} and indeed reproduces super Yang-Mills theory in 10 dimensions. The reason that this construction cannot be repeated for the world line is that in the pure spinor approach the undeformed BRST operator (corresponding to $\Omega^{init}$) above is a derivation of the product of functions on $H$. This not the case in our formalism.

\section*{Acknowledgments}
\label{sec:Aknowledgements}

We profited from discussions with Roberto Bonezzi. I.S. would like to Warren Siegel for helpful discussions. This work was funded by the Excellence Cluster Origins of the DFG under Germany’s Excellence Strategy EXC-2094 390783311. M. G. and I.S. acknowledge the Erwin Schr\"odinger Institute for hospitality while completing this work.

\bibliography{HSmaster}

\end{document}